\documentclass[preprint,aps,nofootinbib,epsfig]{revtex4}
\usepackage{graphicx}

\def\gsim{\begin{array}{c} > \\ \sim \end{array}}
\def\lsim{\begin{array}{c} < \\ \sim \end{array}}

\begin{document}
\draft
\begin{titlepage}
\title{\large \bf Selected topics in Planck-scale physics}
\author{Y. Jack Ng}
\email{yjng@physics.unc.edu}
\address{Institute of Field Physics, Department of Physics and Astronomy,\\
University of North Carolina, Chapel Hill, NC 27599-3255, USA\\}

\bigskip

\begin{abstract}

We review a few topics in Planck-scale physics, with emphasis on
possible manifestations in
relatively low energy.  The selected topics include
quantum fluctuations
of spacetime, their cumulative effects, uncertainties in energy-momentum
measurements, and low energy quantum-gravity phenomenology.
The focus is on quantum-gravity-induced uncertainties in some observable
quantities.
We consider four possible ways to probe Planck-scale physics
experimentally: 1.
looking for energy-dependent spreads in the arrival time of photons
of the same energy from GRBs; 2. examining
spacetime fluctuation-induced phase incoherence of light from
extragalactic sources;
3. detecting spacetime foam with laser-based interferometry techniques;
4. understanding the threshold anomalies in high energy cosmic ray and
gamma ray events.
Some other experiments are briefly discussed.
We show how some physics behind black holes, simple clocks, simple
computers, and the holographic principle is related to Planck-scale physics.
We also discuss a
formulation of the Dirac equation as a difference
equation on a discrete Planck-scale spacetime lattice, 
and a possible interplay
between Planck-scale and Hubble-scale physics encoded in the cosmological
constant (dark energy).

\end{abstract}

\maketitle
\end{titlepage}

\newpage

\begin{verse}
\hspace{2.5in}{Planck was constant.}\\
\hspace{2.8in}{--- {\it The Economist}, March 1st, 2003, p.72}
\end{verse}

\medskip

\section{Introduction}

Planck's constant $\hbar$, Newton's constant $G$, and the speed of light
$c$ can be combined to form the Planck time
$t_P = (\hbar G/c^5)^{1/2} \sim 10^{-44}s$, the
Planck length $l_P = c t_P \sim 10^{-33} cm$, and the
Planck energy $E_P = \hbar / t_P \sim 10^{28} eV$.
Clearly the Planck time is so
short, the Planck length so minuscule, and the Planck energy so high
(in elementary particle physics) that these units are not used casually.
So it
takes a certain amount of foolhardiness to even mention Planck-scale
physics.  Indeed, when Giovanni Amelino-Camelia and the author dared to
give talks on this subject in the Huntsville Workshop 2002, we were
duly recognized and jointly honored with
an award for Physics Exotica by the Executive Organizing Committee
of the Workshop.

Exotic though it may well be, Planck-scale physics has, in recent years, been
garnering wider acceptance in the theoretical physics circle.  The reason is
clear: it is generally believed that
at Planck scale, the quantum aspects of gravity become
manifest.  Only when we understand spacetime at Planck scale can we
properly synthesize quantum mechanics with
general relativity to find the correct theory of quantum gravity.  By now
there are quite a few approaches to quantum gravity.  In addition to the
front runners, string/M theory\cite{string}
and loop quantum gravity\cite{loop}, the list includes
causal sets, dynamical triangulations, causal dynamical triangulations,
twistor theory, non-commutative geometry, supergravity, cellular networks,
approaches based on analogies to condensed matter physics, and
foamy structure of quantum spacetime\cite{hawkellis}.  For an assessment
of these major approaches to quantum gravity
(and references), see Ref.\cite{smolin}.
But in spite of all
the impressive progress these candidate quantum gravity theories have made,
it is probably fair to say that a complete and satisfactory formulation
of the correct theory of quantum gravity is still not yet at hand.

Lacking such a formulation, one can hardly speak of Planck-scale physics with
great confidence.  But by extrapolating the well-known successes of quantum
mechanics and general relativity in low energy, we believe
one can still make
predictions about certain phenomena
involving Planck-scale physics, and check for consistency.
The scope of this Brief Review is quite limited.
It concerns phenomena
at an energy scale much below the Planck energy $E_P$.
It deals only with a few topics in Planck-scale/quantum-gravity
physics, or, more correctly, in the interplay between quantum
mechanics and general relativity --- topics with which the author has some
familiarity.  The focus will mostly be
on the uncertainties in some observable
quantities induced by the synthesis of quantum mechanics and general 
relativity.
Our approach is very conservative.  We make no assumption on the high energy
regime of the ultimate quantum gravity theory, and refrain from
speculating on violations of Lorentz invariance and
systematically modified dispersion relations\footnote{By this, we mean
dispersion relations modified by a term with a coefficient of {\it fixed}
magnitude and sign.  More on this near the end of section VIII.}
which many people believe are
unavoidably induced by quantum gravity.
We want to see how far we can go
without making those assumptions, sensible as they may well be.

The outline of this Brief Review is as follows:
In section II, we introduce the subject by considering
the accuracy with which one can measure a distance or a time interval.
Consistent with our limited objective, the distances
and the time intervals considered are understood to be much larger than
$l_P$ and $t_P$ respectively.  The problem is
tackled from two different angles, by first using a gedanken experiment
of spacetime measurements, and
then by using the holographic principle.  We interpret the resulting
uncertainties in spacetime measurements as due to
quantum fluctuations of spacetime, i.e., the uncertainties in distance/time
measurements are due to fluctuations of the spacetime metric which,
following Wheeler, we will
loosely call the quantum foam or spacetime foam.
Consistency
between the quantum foam picture described in section II and black hole
physics is considered in section III which also gives a discussion of some
connections to limitations to computation and to the accuracy of a
simple clock due to the fuzziness of spacetime .
Fluctuations due to
quantum foam are very minuscule, so they can be detected only if there is
a huge cumulative effect from ``summing'' up the individual fluctuations.  In
the section IV we
consider the cumulative effects of quantum foam.
In section V, we consider how spacetime fluctuations
induce an uncertainty in energy-momentum measurements and possibly modify
dispersion relations.  We also discuss how, in principle,
the speed of light can be used to
test the modified dispersion relation and probe Planck-scale physics.
The next three sections
are devoted to possible ways to do Planck-scale phenomenology.  The trick is
to find measurements which will ``amplify'' small effects of quantum gravity.
In section VI, we consider the possibility of using spacetime foam-induced
phase incoherence of light from distant galaxies
to probe Planck-scale physics (here the ratio of the distance to the galaxies
and the wavelength of light plays the role of the amplifying factor).
In section VII we discuss the the possibility
of measuring the foaminess of spacetime with interferometers (here the
existence of another length scale,
in addition to the Planck length, provided by the frequency of the noise
spectrum plays an important role).
In section VIII
we entertain the idea that energy-momentum uncertainties may be the origin
of threshold anomalies in ultra-high energy cosmic ray and TeV-$\gamma$
events (here the amplifying factor is due to the huge discrepancy between
the energies of the two colliding particles).
We also discuss an alternative proposal involving
``systematic'' deformations of particle dispersion relations.
Concluding remarks are found in the section IX which also
contains
a brief (and necessarily very incomplete) survey of other proposals to
probe Planck-scale physics experimentally.
Two related topics in Planck-scale physics are relegated to appendices.
If spacetime is really discrete at the Planck scale, as suggested by some
approaches to quantum gravity, then one should replace differential equations
describing fundamental interactions by difference equations; in
Appendix A,
we derive the Dirac equation in the form of a difference equation.
In Appendix B, we speculate
on a possible infrared and ultravolet connection,
as a result of an interplay between
Planck-scale and Hubble-scale physics, encoded in the cosmological constant.

\bigskip

\section{\bf Quantum fluctuations of spacetime}

At small scales, spacetime is fuzzy and foamy due to quantum fluctuations.
One manifestion of the fluctuations is in the induced uncertainties in
any distance measurement.  We will derive the uncertainties or fluctuations
by two independent methods.\cite{found,stfoam}  Neither method by itself
is satisfactorily
rigorous, but the fact that they both yield the same result bodes well for
the robustness of the conclusion.
Let us first consider
a gedanken experiment to measure the
distance $l$ between two points.
Following Wigner\cite{wigner}, we can put
a clock at one of the points and a mirror at the other.  By sending a
light signal from the clock to the mirror in a timing experiment, we can
determine the distance.
However, the quantum uncertainty in the positions of
the clock and the mirror introduces an inaccuracy $\delta l$ in the
distance measurement.  Let us concentrate on the clock and denote its mass
by $m$.  Wigner argued that if it has a linear
spread $\delta l$ when the light signal leaves the clock, then its position
spread grows to $\delta l + \hbar l (mc \delta l)^{-1}$
when the light signal returns to the clock, with the minimum at
$\delta l = (\hbar l/mc)^{1/2}$.  Hence one concludes that
\begin{equation}
\delta l^2 \gsim \frac{\hbar l}{mc}.
\label{sw}
\end{equation}
One can supplement this quantum mechanical relation with a limit from
general relativity\cite{ngvan1}.  To see this, let the clock be a
light-clock consisting of two parallel mirrors (each of mass $m/2$), a
distance $d$ apart, between which bounces a beam of light.  For
the uncertainty in distance measurement not to be greater than $\delta l$,
the clock must
tick off time fast enough that $d/c \lsim \delta l /c$.  But $d$, the
size of the clock, must be larger than the Schwarzschild radius $Gm/c^2$ of
the mirrors, for otherwise one cannot read the time registered on the
clock.  From these two requirements, it follows that
\begin{equation}
\delta l \gsim \frac{Gm}{c^2},
\label{ngvan}
\end{equation}
the product
of which with Eq.~(\ref{sw}) yields
\begin{equation}
\delta l \gsim (l l_P^2)^{1/3} = l_P \left(\frac{l}{l_P}\right)^{1/3},
\label{nvd1}
\end{equation}
where $l_P = (\hbar G/c^3)^{1/2}$ is the Planck length.
\footnote{Note that $\delta l$ depends
on both $l_P$ and $l$; it is of mesoscopic nature.  This mesoscopic
scale may best be explored via stochastic semiclassical gravity\cite{Hu}.}
\footnote{We also note that, for (N + 1)-dimensional spacetime, 
Eq.~(\ref{nvd1})
is generalized to read $\delta l \gsim l_P (l/l_P)^{1/N}$.}

A gedanken experiment to measure a time interval
$T$ gives an analogous expression:
\begin{equation}
\delta T \gsim (T t_P^2)^{1/3}.
\label{deltat}
\end{equation}
The spacetime fluctuation
translates into a metric fluctuation over a distance $l$ and a
time interval $T$ given by
\begin{equation}
\delta g_{\mu \nu} \gsim (l_P / l)^{2/3}, \,\,\, (t_P / T)^{2/3},
\label{metricfl}
\end{equation}
respectively.

One can also derive the $\delta l$ result by appealing to the
holographic principle\cite{wbhts} which states that the maximum number of
degrees of
freedom that can be put into a region of space is given by the area of the
region in Planck units.  Consider a
region of space measuring $l \times l \times l$, and imgine partitioning it
into cubes as small as physical laws allow.  With each small cube we
associate one degree of freedom.   If the smallest uncertainty in
measuring a distance $l$ is $\delta l$, in other words, if the fluctuation
in distance $l$ is $\delta l$, then the smallest such cubes have volume
$(\delta l)^3$.  (Otherwise, one could divide $l$ into units each measuring
less than $\delta l$, and by counting the number of such units in $l$, one
would be able to measure $l$ to within an uncertainty smaller than
$\delta l$.)  Thus the maximum number of degrees of freedom,
given by the number of small cubes we can put into the region of
space, is $(l/ \delta l)^3$.  It follows from the holographic principle
that $(l / \delta l)^3 \lsim (l / l_P)^2$, which yields precisely the
same expression for spacetime fluctuation $\delta l$ given by
Eq. (\ref{nvd1}).
In fact one can reverse the argument and argue
that the holographic principle has its origin in the quantum
fluctuations of spacetime.\cite{found,stfoam}
Since the holographic principle is deeply rooted in
black hole physics, this way of deriving spacetime fluctuations is
highly suggestive of the deep connection between quantum foam and black
hole physics.

\bigskip

\section{\bf Clocks, computation, black holes, and Planck-scale physics}

It is interesting that an argument, very similar to that used in
the last section
to deduce the structure of spacetime foam,
can be applied to discuss the precision and
the lifetime of a
clock.\cite{ng}  For a simple clock\footnote{A clock is not simple 
if it is made up of components which are not used to keep time
concurrently.}
of mass $m$, if the
smallest time
interval that it is capable of resolving is $t$ and its total
running time is $T$,
one finds\cite{ng}
\begin{equation}
t^2 \gsim \frac{\hbar T}{mc^2}, \,\,\,\,
t \gsim \frac{Gm}{c^3},
\label{clock1}
\end{equation}
the analogue of Eq.~(\ref{sw}) and Eq.~(\ref{ngvan}) respectively.
One can combine these two expressions to give
\begin{equation}
T/ t^3 \lsim t_P^{-2} = \frac{c^5}{\hbar G},
\label{clock2}
\end{equation}
which relates clock precision
to its lifetime.  (Note that this new expression is just
Eq.~(\ref{deltat}) with $t$ playing the role of $\delta T$.)
For example,
a femtosecond ($10^{-15}$ sec)
precision yields the bound $T \lsim 10^{34}$ years.  However, note that the
bound on $T$ goes down rapidly as $t^3$.

One can translate the above
clock relations into useful expressions for a simple computer.  The fastest
possible processing frequency is obviously given by $t^{-1}$. Thus
we identify $\nu = t^{-1}$ as the clock rate of the computer, i.e., the
number of operations per bit per unit time.\footnote{This corrects a
misidentification of $\nu$ in Ref.\cite{ng}.}
The identification of the
number $I$ of bits of information in the memory space of a simple computer
is subtler.  Since $T/t$ is the maximum number of steps of
information processing, we make the identification $I = T/t$.  Using
the clock relation in Eq. (\ref{clock2})
and the identifications of $\nu$ and $I$
in terms of $t$ and $T$, one gets
\begin{equation}
I \nu^2 \lsim \frac{c^5}{\hbar G} \sim 10^{86} /sec^2.
\label{computer}
\end{equation}
This expression links together our
concepts of information, gravity, and quantum uncertainty.\cite{ng}
We will see below that nature seems
to respect this bound
which, in particular, is saturated for black holes.
For comparison, current laptops
perform about $10^{10}$ operations per sec on $10^{10}$ bits, yielding $I
\nu^2 \sim 10^{10} / sec^2$.

Next let us apply the two (in-)equalities in Eq. (\ref{clock1}) to a black
hole of mass $m$, used as a clock.  It is reasonable to use
the light travel time across the black hole's horizon as the resolution
time of the clock,
i.e., $t \sim \frac{Gm}{c^3} \equiv t_{BH}$, then one immediately finds that
\begin{equation}
T \sim \frac{G^2 m^3}{\hbar c^4} \equiv T_{BH},
\label{Hawking}
\end{equation}
which is just Hawking's black hole
lifetime!  Thus, if we had not known of black hole evaporation, this
remarkable result would have implied that there is a maximum lifetime (of
this magnitude) for a black hole.\cite{ng,barrow}
This is another demonstration of the intimate (if, in this case,
indirect) relationship
between quantum foam and black hole physics.

Now in principle, it is possible to program black holes
to do computations in such a way that the results of the computation
can be read out of the fluctuations in the apparently thermal
Hawking radiation, if black holes indeed evolve in a unitary fashion as we
believe.\cite{Lloyd}  So
imagine that we form
a black hole (of mass $m$) whose initial conditions encode certain
information to be processed.
Then the memory space of the black hole
computer has $I = T_{BH}/t_{BH} \sim (m/m_P)^2$.
This gives the number of bits $I$ as the event horizon
area in Planck units, as expected from the identification\cite{wbhts} of
black hole entropies!  Furthermore, the number of operations per unit time
for a black hole computer is given by $I \nu \sim mc^2/\hbar$, in agreement
with Lloyd's results\cite{Lloyd} for the ultimate physical limits to
computation.

All these results indicate the conceptual interconnections
of the physics underlying simple clocks, simple computers\footnote
{For a quantum computer view of spacetime at the Planck scale, see
Ref.\cite{zizzi}},
black holes, and spacetime foam.  We further note that, for black holes,
the bounds in Eqs. (\ref{clock2}) and (\ref{computer}) are saturated.
Thus one can even claim that black holes are the ultimate simple clocks, and
once they are programmed to do computations, they
are the ultimate computers.  It is curious that although they can be very
massive and large, black holes are basically {\it simple} --- a fact further
supported by the no-hair theorem.  Lastly we want to point out that the
connections between Planck-scale physics and black hole physics discussed in
this section are not unexpected in view of the consistency between spacetime
measurements and the holographic principle discussed in the last section.

\bigskip

\section{Cumulative effects of spacetime fluctuations}

So far we have been discussing a
particular spacetime foam model, motivated by an
elementary consideration of distance measurements and
by the consistency with the holographic
principle.  Let us now introduce a parameter $\alpha
\sim 1$ to specify the different quantum foam models (also called quantum
gravity models).  In terms of the parameter $\alpha$, the distance and metric
uncertainties/fluctuations take the form
\begin{equation}
\delta l \gsim l \left(\frac{l_P}{l}\right)^{\alpha}, \,\,\,\,
\delta g_{\mu \nu} \gsim (l_P / l)^{\alpha}.
\label{genfluct1}
\end{equation}
The standard choice\cite{mis73} of $\alpha$ is $\alpha =
1$; the choice of $\alpha = 2/3$ discussed above appears\cite{found,ng}
to be consistent with the holographic
principle and black hole physics and will be called the holography model;
$\alpha = 1/2$ corresponds to the random-walk model
found in the literature\cite{AC,dio89}.
Though much of our discussion below is applicable to the general case,
we will
use these three cases as examples of the quantum gravity models.
Note that all the three quantum gravity models predict a very small
distance uncertainty: e.g.,
even on the size of the whole observable universe ($\sim 10^{10}$
ligh-years), Eq. (\ref{genfluct1}) yields a fluctuation of only about
$10^{-33}$ cm, $10^{-13}$ cm, and $10^{-2}$ cm for $\alpha = 1, 2/3$ and
$1/2$ respectively.

Let us now examine the cumulative effects\cite{NCvD}
of spacetime fluctuations over
a large distance.
Consider a distance $L$ (which will denote the distance
between extragalactic sources and the telescope in section VI), and
divide it into $L/ \lambda$ equal
parts each of which has length $\lambda$ (which, for the discussion in
section VI, will {\it naturally}
denote the wavelength of the observed light from the distant source).
If we start with $\delta \lambda$ from each part, the question is how do
the $L/ \lambda$ parts
add up to $\delta L$ for the whole distance $L$.  In other words, we want
to find
the cumulative factor $\mathcal{C}_{\alpha}$ defined by
\begin{equation}
\delta L = \mathcal{C}_{\alpha}\, \delta \lambda,
\label{cf1/2.1}
\end{equation}
Since $\delta L \sim l_P (L/l_P)^{1 - \alpha}$ and
$\delta \lambda \sim l_P (\lambda/l_P)^{1 - \alpha}$, the result is
\begin{equation}
\mathcal{C}_{\alpha} = \left({L \over \lambda}\right)^{1 - \alpha},
\label{cf}
\end{equation}
in particular,
\begin{equation}
\mathcal{C}_{\alpha=1/2} = (L/ \lambda)^{1/2}, \hspace{.1in}
\mathcal{C}_{\alpha=2/3} = (L/ \lambda)^{1/3}, \hspace{.1in}
\mathcal{C}_{\alpha=1} = (L/ \lambda)^0 = 1,
\label{cfs}
\end{equation}
for the random walk $\alpha = 1/2$ case,
the holography $\alpha = 2/3$ case and the ``standard" $\alpha = 1$ case
respectively.  Note that $\mathcal{C}_{\alpha=1} = 1$ is
{\it independent} of $L$.
Strange as it may seem, the result is not
unreasonable if we recall, for the ``standard'' model,
$\delta l \gtrsim l_P$,
independent of $l$.  The crucial point to remember is that, for all
the quantum gravity models, {\it none} of the cumulative
factors is linear in $(L/\lambda)$, i.e.,
\begin{equation}
{\delta L \over \delta \lambda} \neq {L \over \lambda}.
\label{nonlinear}
\end{equation}
The reason for this is obvious: the $\delta \lambda$'s from the $L/ \lambda$
parts in $L$ do {\it not} add
coherently.\footnote{To gain insight into the process,
consider the $\alpha = 1/2$
random-walk model of quantum gravity,
and for simplicity, assume
that $\delta \lambda$ takes on only two values, viz. $\pm l_P
(\lambda/l_P)^{1/2}$,
with equal probability
(instead of, say, a Gaussian distribution about zero, which is more likely).
If the fluctuations from the different
segments are all of the same sign, then together they contribute
$\pm l_P (\lambda /l_P)^{1/2} \times (L/\lambda)$ to $\delta L$.
But both these two cases, yielding a linear $L$-dependence
for $\delta L$, are extremely
unlikely (each having a probablity of $1/2^{L/\lambda} \ll 1$ for
$(L/\lambda) \gg 1$.)
For this one-dimensional random walk involving $L/\lambda$ steps of
equal size ($\delta \lambda$), each step moving right or left
(corresponding to $+$ or $-$ sign) with equal
probability, the result is well-known: the cumulative fluctuation is
given by $\delta L \sim \delta \lambda \times (L/\lambda)^{1/2}$ which
is $l_P (L/\l_P)^{1/2}$ as expected for consistency.}
In fact, according to Eq.~(\ref{cf}),
the cumulative effects are linear in $L/\lambda$
only for the physically unacceptable case of
$\alpha = 0$ for which $\delta l \sim l$.
To obtain the correct cumulative factor (given by Eq.~(\ref{cf}))
from what we may inadvertently think it is, viz.,
$(L/\lambda)$ (independent of $\alpha$), we have to put in the
{\it correction factor} $(\lambda/L)^{\alpha}$.

\bigskip

\section{\bf Energy-momentum uncertainties}

Just as there are uncertainties in distance and time interval measurements,
there are uncertainties in energy-momentum measurements.  Both types
of uncertainties\cite{ngvan1,nvD1}
come from the same source, viz., quantum fluctuations of space-time
metrics\cite{Ford} giving rise to space-time foam.
Imagine sending
a particle of momentum $p$ to probe a certain structure of spatial extent 
$l$ so that $p \sim \frac{\hbar}{l}$.
Consider the coupling of the metric to the energy-momentum tensor of the
particle,
$(g_{\mu \nu} + \delta g_{\mu \nu}) t^{\mu \nu} = g_{\mu \nu} (t^{\mu \nu}
+ \delta t^{\mu \nu})$,
where we have noted that the uncertainty in $g_{\mu \nu}$ can
be translated into
an uncertainty in $t_{\mu \nu}$.
Eq. (\ref{genfluct1}) for $\delta g_{\mu \nu}$ can
now be used to give
\begin{equation}
\delta p = \beta p \left(\frac{p}{m_P c}\right)^{\alpha},
\label{dp}
\end{equation}
where $\beta \sim 1$
and $m_P$ is the Planck mass.\footnote{Alternatively, one can simply use
$p \sim \frac{\hbar}{l}$ (so that $\delta p \sim (\hbar/l^2) \delta l$)
in conjunction with $\delta l \gtrsim l (l_P/l)^{\alpha}$.}
Another way to derive the momentum uncertainty is to
regard $\delta p$ as the uncertainty of the momentum operator $p =
-i \hbar \partial / \partial x$, associated with
$\delta x = x (l_P / x)^{\alpha}$.\cite{ngvan1,nvD1}
The corresponding statement for energy uncertainties is
\begin{equation}
\delta E = \gamma E \left(\frac{E}{E_P}\right)^{\alpha},
\label{de}
\end{equation}
with $\gamma \sim 1$.
Note that the energy-momentum uncertainty is actually fixed by dimensional
analysis,
once the uncertainty in the metric is given by Eq. (\ref{genfluct1}).
We emphasize that all the uncertainties take on $\pm$ sign with equal
probability (most likely, a Gaussian distribution about zero).

What is the time scale at which these fluctuations occur?  Returning to
the gedanken experiment in distance $l$ measurement discussed
in section II, we may be tempted to
conclude that the $\delta l$ fluctuation occurs at a time scale given by
$l/c$, independent of the Planck-scale.  Accordingly, the energy fluctuation
$\delta E$ occurs at a time scale $\hbar / E$.  However, this argument is
by no means convincing.  While it does take $\sim l/c$ amount to time
to make one measurement of the distance, to measure the fluctuations of
the distance we need to take more than one measurement.  We are led to
ask how quickly we can make a succession of measurements of the distance 
and perhaps then we would argue that it is the time separation between
the different measurements that is the relevent time scale at which the
fluctuations occur.  The question of which
is the correct time scale to use has yet to be settled.

Energy-momentum uncertainties affect both the energy-momentum conservation
laws and the dispersion relations.  Energy-momentum is
conserved up to energy-momentum uncertainties due to quantum foam effects,
i.e., $\Sigma (p_i^{\mu} + \delta p_i^{\mu}$)
is conserved with $p_i^{\mu}$ being the average values.  On the other hand
the dispersion relation is now generalized to read
\begin{equation}
E^2 - p^2 - \epsilon p^2 \left({p \over E_P}\right)^{\alpha} = m^2,
\label{moddisp}
\end{equation}
for high energies with $E \gg m$.  Here and hereafter, unless clarity
demands otherwise, we set the speed of light $c$, as well as Planck's
constant $\hbar$, equal to unity.  For $\epsilon \neq 0$, one can interpret
the additonal term in the dispersion relation as an additional contribution
to the mass which is energy-dependent.
{\it A priori}
we expect $\epsilon \sim 1$ and is independent of $\beta$ and $\gamma$.  But
due to our ignorance of quantum gravity, we cannot make any definite
statements.  It is possible that $\epsilon \approx 2 (\beta - \gamma)$, which
would be the case if the dispersion relation is given by
$(E + \delta E)^2 - (p + \delta p)^2 = m^2$.  Another possibility is that
$\epsilon \approx 0$, which would be the case if the
usual dispersion relation holds for the average $E$ and $p$.  One can reach
the latter conclusion if one appeals to the
van-Dam-Veltman-Zakharov discontinuity theorem\cite{vdv} which states that
the theory for an exactly massless graviton is different from that for an
extremely light graviton.  If graviton is indeed the quantum mediator of
gravitational interactions (in a Minkowskian spacetime),
according to Ref.\cite{vdv}, only the theory for
an exactly massless graviton can
explain Einstein's three tests of general relativity.  To the extent that
the vDVZ theorem is correct, one infers that
the dispersion relation for gravitons is unaffected
by quantum fluctuations of spacetime, i.e., $ \epsilon =
0$, so that the graviton does not have an
energy-dependent effective mass and remains exactly massless.
Then it is not hard to imagine that the dispersion relation for other
particles, especially the massless particles like the photon,
is also not affected.

But if $\epsilon \neq 0$,
the modified dispersion relation discussed above has an interesting
consequence for the speed of light.  Applying Eq.~(\ref{moddisp})
to the massless photon yields
\begin{equation}
E^2 \simeq c^2p^2 + \epsilon E^2 \left(\frac{E}{E_P}\right)^{\alpha},
\label{gamd}
\end{equation}
where we have restored the factor of c .
The speed of (massless) photon
\begin{equation}
v = \frac{\partial E}{\partial p} \simeq c
\left( 1 + \epsilon \frac{1 + \alpha}{2}
\frac{E^{\alpha}}{E_P^{\alpha}}\right),
\label{gams}
\end{equation}
becomes energy-dependent if $\epsilon \neq 0$, and it fluctuates around c.
This fluctuating speed of light would {\it seem} to 
yield\cite{NLOvD} an energy-dependent spread
in the arrival times of photons of the {\it same} energy $E$ given by
$\delta t \sim |\epsilon|t (E/E_P)^{\alpha}$,
where $t$ is the average overall time of travel from the photon source, say,
a gamma ray burster, for which both $t$ and $E$ are relatively large.
Furthermore, the modified energy-momentum dispersion relation would seem to
predict
time-of-flight differences between simultaneously-emitted photons of
different energies, $E_1$ and $E_2$, given by
\begin{equation}
\delta t \simeq \epsilon t \frac{1 + \alpha}{2}
\frac{E_1^{\alpha} - E_2^{\alpha}}{E_P^{\alpha}}.
\label{tdiff}
\end{equation}
An upper bound\cite{Ellis,ACP} on the absolute value of $\epsilon$ 
could then be obtained
from the observation\cite{tevob} of simultaneous (within experimental
uncertainty of $\delta t \leq 200$ sec) arrival of 1-TeV and 2-TeV
$\gamma$-rays from Mk 421 which is believed to be $\sim 143$ Mpc away from
the Earth.   
But these results for the spread of arrival times of photons are
{\it not} correct, because we have inadvertently forgotten to put in the 
necessary correction factors discussed in the last section.  For the 
spread in arrival time of the photons of the same energy, taking into
account the correction factor $(\lambda /L)^{\alpha} \sim 
(\hbar /Et)^{\alpha}$, we get a much smaller $\delta t \sim 
t^{1 - \alpha} t_P^{\alpha}$.  We will witness the importance of the 
correction factor again in the next section.  But note that, if 
$\epsilon$ is constant (instead of fluctuating about zero), then 
Eq.~(\ref{tdiff}) does give the correct time-of-flight differences.  See 
sections VIII and IX.

\bigskip

\section{\bf Induced
phase incoherence of light from galaxies}

Recently Lieu and
Hillman\cite{lie02b,lie03} and then Ragazzoni, Turatto, and Gaessler
\cite{rag03} proposed a technique
that has hitherto been overlooked to directly test the Planck scale
fluctuations.  They argued that these fluctuations can cumulatively lead
to a complete loss of phase for radiations that have propagated a
sufficiently large distance and they searched for patterns of images
of very distant galaxies gathered by
telescopes that should not be present if prevailing notions of spacetime
quantum is correct.  In this section, we\cite{NCvD} critically examine their
very interesting idea.

Consider the phase
behavior of light with wavelength $\lambda$ received from a celestial
optical source located at a distance $L$ away.  During the propagation
time $T = L/v_g$ where $v_g$ is the group velocity of propagation, the
phase has advanced by the amount
\begin{equation}
\phi = 2 \pi {v_p T\over \lambda} = 2 \pi {v_p\over v_g}{L\over \lambda},
\label{phase1}
\end{equation}
where $v_p$ is the phase velocity of the light wave.  This phase fluctuates
randomly according to
\begin{eqnarray}
\delta \phi &=& 2 \pi {L\over \lambda}\, \delta\!\!\left({v_p\over v_g}\right)
   + 2 \pi {v_p \over v_g} L\, \delta\!\!\left({1\over \lambda}\right)
    + 2 \pi {v_p \over v_g}\, {1\over \lambda}\, \delta\!L\nonumber\\
&=& \delta \phi_1 + \delta \phi_2 + \delta \phi_3,
\label{phase2}
\end{eqnarray}
with $\delta \phi_i$ denoting the three successive terms
in $\delta \phi$ ($i = 1, 2, 3$).
Note that the various expressions for $\delta \phi_i$ have been obtained
by simple-minded straightforward algebra.  The quantum gravitational and
statistical nature of the fluctuations have yet to be properly incorporated,
i.e., we still have to fold in the correction factors for the cumulative
effects, as discussed in section IV.

For $\delta \phi_1$, applying Eq.~(\ref{moddisp}) with $m = 0$ for photon,
and recalling that $v_p = E/p$ and $v_g = dE / dp$, we obtain
\begin{equation}
\delta\! \left( {v_p \over v_g}\right) \sim \left({E\over E_P}
\right)^{\alpha} = \left({l_P \over \lambda}\right)^{\alpha},
\label{deltav}
\end{equation}
where we have used $E/E_P = l_P / \lambda$ and $\epsilon \sim 1$.
Putting in the correction
factor $(\lambda/L)^{\alpha}$, we obtain
\begin{equation}
\delta \phi_1 \sim  2 \pi {L\over \lambda} \left({l_P\over
\lambda}\right)^{\alpha} \left({\lambda \over L}\right)^{\alpha}
= 2 \pi {l_P^{\alpha} L^{1 - \alpha} \over \lambda}.
\label{delphi}
\end{equation}
For $\delta \phi_2$, it suffices to approximate $v_p/v_g$ by 1.  Recalling
that $\delta \lambda \sim l_P (\lambda /l_P)^{1 - \alpha}$ and the need to
put in the correction factor $(\lambda/L)^{\alpha}$, we get
$\delta \phi_2 \sim \delta \phi_1$.
For $\delta \phi_3$, also approximating $v_p/v_g$ by 1,
using $\delta L \sim l_P (L / l_P)^{1 - \alpha}$, and noting that there is
no need for a correction factor for this term,
we immediately find $\delta \phi_3 \sim \delta \phi_1$.
Since all the three $\delta \phi_i$'s
are of the same order of magnitude, we conclude that
\begin{equation}
\delta \phi = 2 \pi a {l_P^{\alpha} L^{1 - \alpha} \over \lambda}
\sim 2 \pi {\delta L \over \lambda},
\label{totphi}
\end{equation}
where $a \sim 1$.  In passing, we note that, since $\delta \phi_1$ involves
energy-momentum fluctuations whereas both $\delta \phi_2$ and
$\delta \phi_3$ involve distance fluctuations, the fact that they all make
contributions of the same order of magnitude can be taken as a sign of
consistency between Eq.~(\ref{genfluct1}), Eq.~(\ref{dp}) and
Eq.~(\ref{de}).

In stellar interferometry, following Lieu and Hillman's\cite{lie02b,lie03}
reasoning, for light
waves from an astronomical source incident upon two reflectors
(within a terrestrial telescope) to
subsequently converge to form interference fringes, it is necessary
that $\delta \phi \lesssim 2 \pi$.
But the analysis of the principles of interferometry of distant
{\it incoherent} astronomical ``point'' sources can be tricky.  The
local
spatial coherence across an interferometer's aperature for photons from a
distant point source (i.e., plane waves) is a reflection of the fact that
all photons have the same resultant phase differences {\it across the
interferometer}.  However, as Lieu and Hillman pointed out, this local
coherence can be lost if there is an intervening medium such as a
turbulent plasma or spacetime foam capable of introducing small changes
into the ``effective'' phases of the photon stream falling on the
interferometer.  Such spacetime foam-induced
phase differences are themselves incoherent
and therefore must be treated with the {\it correct cumulative factors}
$\mathcal{C}_{\alpha}$
appropriate for the quantum gravity model under consideration.

First we note that since
the cumulative factor for the ``standard'' model of quantum gravity
(for which $\alpha = 1$) is 1, i.e., there is no cumulative effect,
obviously Lieu and Hillman's proposed approach
cannot be used to
rule out (or confirm) the $\alpha = 1$ model.
To rule out a certain model with $\alpha < 1$, the strategy is to look for
unexpected interference fringes for which the phase coherence of light from
the distant sources should have been lost
(i.e., $\delta \phi \gtrsim 2 \pi$)
for that value of $\alpha$ according to theoretical calculations.
Consider the example cited by Lieu and Hillman\cite{lie03},
involving the clearly visible
Airy rings in an observation of the active
galaxy PKS1413+135 ($L$ = 1.216 Gpc) by the HST at $\lambda =1.6 \mu m$
wavelength\cite{per02}.  For this example, Eq.~(\ref{totphi}) yields
$\delta \phi \sim 10 \times 2 \pi a$ for the random walk
$\alpha = 1/2$ model and
$\delta \phi \sim 10^{-9} \times 2 \pi a$ for the holography
$\alpha = 2/3$ model.
Since we expect $a \sim 1$, the observation of Airy rings in this case would
seem to marginally, if at all, rule out the random walk model.
On the other hand, the holography model is
obviously not ruled out.  This finding contradicts the
conclusion reached recently by Lieu and Hillman\cite{lie03} who argued that
the HST detection of Airy rings from PKS1413+135 has ruled out a majority of
modern models of quantum gravity, including the ``standard'' $\alpha = 1$
model.  (Earlier, Lieu and Hillman\cite{lie02b}
had claimed to have ruled out the $\alpha = 2/3$ model by
noticing that interference effects were clearly seen in
the Infra-red Optical Telescope Array\cite{van02} at
$\lambda = 2.2 \mu m$ light from the star S Ser which
is $\sim 1$ kpc away.)  The resolution of this disagreement lies in the
fact that Lieu and Hillman
neglected to take into account the correction factor in
estimating the cumulative effects of spacetime foam.  This neglect resulted
in their overestimate of the cumulative effects by a factor
$(L/\lambda)^{\alpha}$: for the case of PKS1413+135, by $10^{20}$
and $10^{30}$ for $\alpha = 2/3, 1$ respectively.  Subsequent work by
Ragazzoni et al.\cite{rag03} contains the same error
of assuming that the cumulative factor is $(L/\lambda)$ rather than the
correct factor $(L/\lambda)^{1 - \alpha}$.  Their claim that the
$\alpha = 2/3$ model and the $\alpha = 1$ model are ruled out is also
far from being justified.  We note that
Coule\cite{cou03} has independently pointed
out that ``Planck scale is still safe from stellar images'' using another
argument.

\bigskip

\section{\bf Detecting quantum foam with
interferometers}

As pointed out recently~\cite{AC,found}, modern
gravitational-wave interferometers, through future refinements, may reach
displacement noise level low enough to test a subset of the space-time
foam models.
To see this, in any distance measurement that involves
a time interval $\tau$, we note that there is a
minute uncertainty
\begin{equation}
\sigma \sim l_P^{\alpha} (c\tau)^{1 - \alpha}.
\label{noise}
\end{equation}
This uncertainty
manifests itself as a displacement noise (in addition to other sources of
noises) that infests the interferometers.
Modern gravitational-wave interferometers are sensitive to changes in
distances to an accuracy of $\sim 10^{-18}$ m or better.  True, this
extraordinary sensitivity is still no where near the Planck length.  But
what really counts is whether the length scale characteristic of the
associated noise of quantum foam at the frequency of the
interferometer bandwidth is comparable to the sensitivity level of the
interferometer.  For an interferometer with bandwidth centered at frequency
$f$, the relevant length scale characteristic of the noise due to
spacetime foam is given by $l_P^{\alpha} (c/f)^{1 - \alpha}$.  Interestingly,
within certain range of frequencies, the experimental limits are comparable
to the theoretical predictions for some of the quantum gravity models.

One can analyse the displacement noise in terms of the associated
displacement amplitude spectral density $S(f)$ of frequency $f$.  For a
frequency-band limited from below by the time of observation $t$, $\sigma$
is given in terms of $S(f)$ by\cite{radeka}
\begin{equation}
\sigma^2 = \int_{1/t}^{f_{max}}[S(f)]^2 df.
\label{spden}
\end{equation}
Now we can easily check that, for the displacement noise given by Eq.
(\ref{noise}), the
associated $S(f)$ is
\begin{equation}
S(f) \sim c^{1 - \alpha} l_P^{\alpha} f^{ \alpha - {3 \over 2}}.
\label{SD}
\end{equation}

By comparing the spectral density with the existing observed noise
level\cite{abram}
of $3 \times 10^{-17} {\rm cm-Hz}^{-1/2}$ near 450 Hz, the lowest noise level
reached by the Caltech 40-meter interferometer, we obtain the bound
$l_P \lsim 10^{-15}, 10^{-27}$ and $10^{-38}$ cm
for the quantum gravity models given by
$\alpha = 1, 2/3$ and $1/2$ respectively.
The ``advanced phase'' of LIGO\cite{abram2} is
expected to achieve a displacement noise level of less than $10^{-20}
{\rm mHz}^{-1/2}$ near 100 Hz, and this would probe $l_P$ down to
$10^{-17}, 10^{-31}$ and $10^{-43}$ cm for $\alpha = 1, 2/3$ and
$1/2$ respectively.  This analysis seems to suggest that the random walk
$\alpha = 1/2$ model is already ruled out.  But more excitedly, modern
gravitational-wave interferometers appear to be within striking distance
of testing the holography ($\alpha = 2/3$) quantum gravity model.
Since $S(f) \sim f^{-5/6}$ for this model (see Eq. (\ref{SD})), we can
optimize the performance of an interferometer at low frequencies.  As
lower frequency detection is possible only in space, one might think that
the planned LISA\cite{Danzmann} is more suitable for our purpose.
Unfortunately, LISA loses more due to its greater arm length than what
it gains by going to lower frequencies.\footnote{As noted by Amelino-Camelia
\cite{AC00}, the relevant quantity is the strain noise power spectrum which
is given by displacement spectral density divided by the square of the arm
length.}

We also note that the correlation length of quantum gravity
fluctuation noise is extremely short as the characteristic scale is the
Planck length.  Thus it can easily be distinguished from other noises
because of this lack of correlation.  In this regard, it may
be useful for quantum gravity studies to have two nearby interferometers.

So far we have concentrated on the observation along the propagation
direction of light in the interferometer.  A matter of concern is the
effect of the beam size in the transverse direction.\footnote{This part of
the discussion is based on private communications with G. Amelino-Camelia
and R. Weiss.}  Implicit
in the discussion above is the assumption that spacetime in between the
mirrors in the interferometer fluctuates coherently for all the photons in
the beam.  But the large beam size in LIGO (compared to the Planck scale)
makes such coherence unlikely.  Thus a small beam interferometer of
comparable power and phase sensibility would definitely be
much more sensitive to the predicted effects of quantum gravity.
Obviously building such a dedicated interferometer would be very valuable.
But we cannot emphasize enough the importance of LIGO
achieving its best noise limit which, even in the form of negative results,
will still be of utmost interests to us.

Finally we mention that there have been suggestions to use atom
interferometers
and optical interferometers
\cite{Per,stfoam,group,Crow} (with minimal beam size effects)
to look for effects of spacetime fluctuations.

\bigskip

\section{Energy-momentum uncertainties and the UHECR events}

The universe appears to be more transparent to the ultra-high energy
cosmic rays (UHECRs)\cite{CRexpt} and multi-TeV $\gamma$-rays\cite{GRexpt}
than expected.  Theoretically one expects the
UHECRs to interact with the Cosmic Microwave Background
Radiation and produce pions, and the TeV photons to
interact with the Far Infra Red Background (FIRB)
photons and produce electron-positron
pairs.  These interactions above the respective threshold energies
should make observations of UHECRs
with $E > 5 {\cdot} 10^{19}$eV (the GZK limit)\cite{GZK} or of
gamma-rays with $E > 10$TeV\cite{NGS} from distant sources
unlikely.  Still UHECRs above the GZK limit and, with much less
data and some uncertainties about the FIRB,
Mk501 photons with energies up to 24 TeV have been observed.
In this section, we attempt to explain the
(well-established) UHECR paradox and the (not so-well-established)
TeV-$\gamma$ puzzle, by arguing\cite{NLOvD}
that energy-momentum uncertainties
due to quantum gravity (significant only for high energy particles
like the UHECRs and TeV-$\gamma$-rays),
too small to be detected in low-energy
regime, can
affect particle kinematics so as to raise or even eliminate the
energy thresholds, thereby explaining the threshold anomalies, in
these two reactions.\footnote{Unfortunately, we have nothing useful
to say about the origins of these energetic particles per se.}
(For similar or related approaches, see Ref.\cite{otherexpl}.)

Relevant to the discussion of the UHECR events
and the TeV-$\gamma$ events is the scattering process in which an energetic
particle
of energy $E_1$ and momentum $\mathbf{p}_1$ collides head-on with a soft
photon of
energy $\omega$ in the production of two energetic particles with
energy
$E_2$, $E_3$ and momentum $\mathbf{p}_2$, $\mathbf{p}_3$.
After taking into account energy-momentum uncertainties,
the conservation laws demand
\begin{equation}
E_1 + \delta E_1 + \omega = E_2 + \delta E_2 +E_3 + \delta E_3,
\label{mec}
\end{equation}
and
\begin{equation}
p_1 + \delta p_1 - \omega = p_2 + \delta p_2 + p_3 + \delta p_3,
\label{mmc}
\end{equation}
where $\delta E_i$ and $\delta p_i$
($i = 1, 2, 3$) are given by Eqs. (\ref{de}) and (\ref{dp}),
\begin{equation}
\delta E_i = \gamma_i E_i \left(\frac{E_i}{E_P}\right)^{\alpha},\,\,\,\,
\delta p_i = \beta_i p_i \left(\frac{p_i}{m_P c}\right)^{\alpha},
\label{dep}
\end{equation}
and we have omitted $\delta \omega$, the contribution from the uncertainty
of $\omega$, because $\omega$ is small.\footnote{We should mention that
we have not found the proper (possibly nonlinear) transformations of
the energy-momentum uncertainties between different reference frames.
Therefore we apply the results only in the frame in which we do the
observations.}

Combining Eq.~(\ref{dep}) with the modified dispersion
relations\footnote
{The suggestion that the dispersion relation may be modified by quantum
gravity first appeared in Ref.\cite{IJMP}.}
Eq. (\ref{moddisp}) for the incoming energetic particle ($i=1$)and the
two outgoing particles ($i=2,3$),
\begin{equation}
E_i^2 - p_i^2 - \epsilon_i p_i^2 \left({p_i \over E_P}\right)^{\alpha}
= m_i^2,
\label{moddispi}
\end{equation}
we obtain the threshold energy equation
\begin{equation}
E_{th} = p_0 + \tilde{\eta} {1\over 4\omega}
   {E_{th}^{2 + \alpha} \over E_P^{\alpha}},
\label{Eth}
\end{equation}
where
\begin{equation}
p_0 \equiv \frac{(m_2 + m_3)^2 - m_1^2}{4 \omega}
\label{p0}
\end{equation}
is the (ordinary) threshold energy if there were no energy-momentum
uncertainties, and
\begin{equation}
\tilde{\eta} \equiv \eta_1 - \frac{\eta_2 m_2^{1 +\alpha} +
        \eta_3 m_3^{1 + \alpha}}{(m_2 + m_3)^{1 + \alpha}},
\label{teta}
\end{equation}
with
\begin{equation}
\eta_i \equiv 2\beta_i - 2\gamma_i - \epsilon_i.
\label{eta}
\end{equation}
Note that, in Eq.~(\ref{Eth}), the quantum gravity correction term is
enhanced by the fact that $\omega$ is so small (compared to $p_0$).

Given that all the $\beta_i$'s, the $\gamma_i$'s and the $\epsilon_i$'s are
of order 1 and can be $\pm$, $\tilde{\eta}$ can be $\pm$ (taking on some
unknown Gaussian distribution about zero),
but it cannot be much bigger than 1 in magnitude.
For positive $\tilde{\eta}$, $E_{th}$ is greater than $p_0$.
The threshold energy increases with $\tilde{\eta}$ to ${3\over 2} p_0$ at
$\tilde{\eta} = \tilde{\eta}_{max}$, beyond which
there is no (real) physical solution to Eq.~(\ref{Eth}) (i.e.,
$E_{th}$ becomes complex) and we interpret this as {\it evading} the
threshold cut.\cite{NLOvD}
The cutoff $\tilde{\eta}_{max}$ is very small: $\tilde{\eta}_{max}
\sim 10^{-14}, 10^{-17}$ for $\alpha = 1, 2/3$ respectively for UHECRs; it
is more modest for TeV-$\gamma$-rays: $\tilde{\eta}_{max} \sim 1, 10^{-5}$
for
$\alpha = 1, 2/3$ respectively.  Thus, energy-momentum uncertainties due to
quantum gravity, too small to be detected in low-energy regime, can (in
principle) affect particle kinematics so as to raise or even eliminate
energy thresholds.  Can this be the solution to the UHECR and multi-TeV
$\gamma$-ray threshold anomaly puzzles?

On the other hand, for negative $\tilde{\eta}$, the threshold
energy is less than $p_0$, i.e., a negative $\tilde{\eta}$ {\it lowers}
the threshold energy.\cite{ACNvD,Aloisio2,Huntsville}
For the case of multi-TeV $\gamma$-rays,
$\tilde{\eta} \sim -1$ yields $E_{th}
\sim 0.9 p_0$ for $\alpha = 1$ model.  The situation for the case of
UHECRs is more interesting: $\tilde{\eta} \sim -1$ gives $E_{th} \sim
10^{15}$eV for $\alpha = 1$.\cite{Aloisio2,Huntsville}  Can this be the
explanation of the opening
up of the ``precocious'' threshold in the ``knee'' region?

\begin{figure}
\centering
\includegraphics[height=3in]{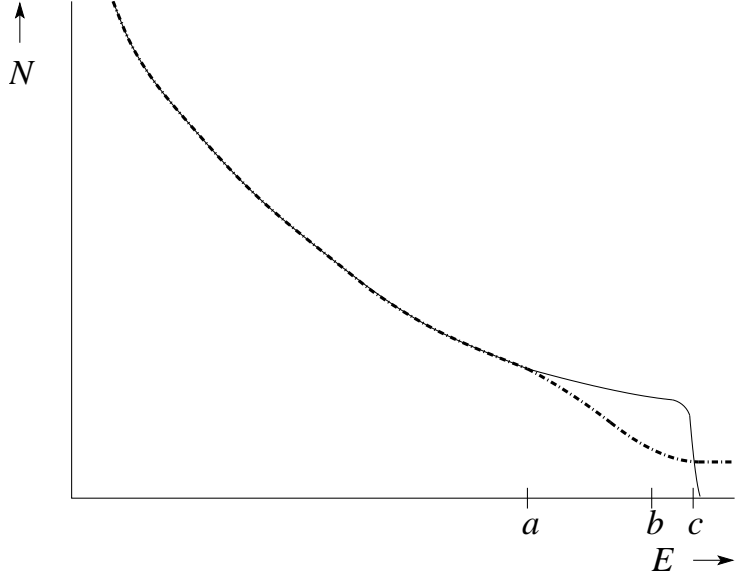}
\caption{Schematic plot of the number N of UHECRs versus energy E.  The solid 
curve refers to the case of ordinary threshold energy $E_{th} = p_0$.  The
dashed-dotted curve refers to the case of the threshold energy given by 
Eq.~(\ref{Eth}).  The ``knee'' region is indicated by ``a'', the ``ankle''
region by ``b'', and the GZK limit by ``c''.}
\label{kat}
\end{figure}

It is far too early to call this a success.  But an
optimistic assessment, indicated by the 
schematic plot in Figure~\ref{kat}
of the number of UHECRs versus energy, invites
one to wonder if the ``ankle'' region is also ``explained''.
There is a similar plot for the TeV-$\gamma$ events.

However, before we get carried away, we should
be aware that there are at least
two problems that confront this particular proposal to solve the two
astrophysical puzzles (besides the obvious problem of making the analysis
quantative).  First of all, there is no
guarantee that $\tilde{\eta}$ is not too small;
if, e.g., $\mid \tilde{\eta} \mid \leq \tilde{\eta}_{max}$, then the
problems are not ameliorated.  But even if $\tilde{\eta}$ spreads over a
large enough range of $\pm$ values, there is still a potentially
serious barrier to overcome.  Let us concentrate on the case of TeV-
$\gamma$-rays.  In the above discussion\cite{ACNvD}, we have analyzed
a single photon-photon collision, focusing on the kinematic
requirements for electron-positron pair production.
For a Mk501 photon with energy of some 10 or 20 TeV,
there are many opportunities to collide with soft photons
with energy suitable for pair production to occur.\footnote{The mean
free path is much shorter than the distance between the source
and the Earth.}
Thus one expects
that even a small probability of producing an electron-positron
pair in a single collision might be sufficient
to lead to the disappearance of the MK501 hard photon before
reaching our detectors. The probability is small in a
single collision with a soft background photon, but
the fact that there are, during the long journey,
many such pair-production opportunities
renders it likely that in one of the many collisions
the hard photon would indeed disappear into an electron-positron
pair.   Completely analogous arguments apply to
the analysis of ultra-high-energy cosmic rays.
While this does not exclude altogether the idea
that energy-momentum fluctuation effects might be responsible for the
threshold anomalies, it does demonstrate the need for further study of this
particular scenario.

So far we have considered the physics purely based 
on quantum-gravity-induced uncertainties.  But now we realize that
the potential difficulties encountered by the above proposal to solve the
UHECR and multi-TeV $\gamma$-ray threshold puzzles can in fact be overcome
if $\epsilon$ in the modified dispersion relation takes on a {\it fixed}
(large enough) {\it negative}\footnote{The positive sign is rejected as
required by matter stability.  More on matter (in)stability later.}
value
\begin{equation}
E_i^2 - p_i^2 - \epsilon p_i^2 \left({p_i \over E_P}\right)^{\alpha}
= m_i^2,\,\,\, \epsilon < 0,
\label{systmod}
\end{equation}
and the ordinary energy-momentum conservation
holds (i.e., $\beta = \gamma = 0$).
Unlike the above proposal which involves fundamental uncertainties in
some observable quantities like energy-momentum, this
proposal\footnote{Chronologically, this proposal was put forth earlier.
See, e.g., Ref.\cite{ACP}.}
systematically shifts the threshold energies.  In the terminology of
Ref.\cite{ACNvD}, the above approach is said to involve a non-systematic
effect of quantum gravity while the latter (different) approach
involves a systematic quantum
gravity effect.  It has been argued\footnote{For example, in the framework
of loop quantum gravity.  The ``modified special relativity'' or
``doubly special relativity'' approach\cite{dsr} provides another example
where
the dispersion relation is systmatically modified, though the expressions
for the conservation of energy-momentum are in general also modified
in this approach.}
that quantum
gravity-induced deviations from ordinary Lorentz invariance
at Planck scale might lead to
such a systematic deformation of the dispersion relation.
In that case, both systematic
and non-systematic effects may be present.
It is therefore important that we
understand how systematic effects and nonsystematic effects
can combine in physical contexts such as the ones
pertaining to the paradoxes we have considered.\cite{ACNvD}
A systematic effect may well raise the threshold for
a particle-production process, but in the presence of an
accompanying non-systematic effect, this increase of the threshold
will have to be interpreted only in a statistical sense:
processes with energetics below the new higher threshold can still
(with however small probability) occur.

In addition to the potential difficulties discussed above, there is
another potential problem for the approach involving
``non-systematic'' effects of quantum gravity.  There is the question
of matter (in)stability\cite{stability} in connection with this
approach because quantum fluctuations in dispersion relations
Eq.~(\ref{moddispi}) can {\it lower} as well as raise the reaction thresholds.
This problem may force us to entertain 
one or a combination of the following possibilities: (1) The
fluctuations of the energy-momentum of a particle are not completely
uncorrelated (e.g, the fluctuating coefficients $\beta$, $\gamma$, and
$\epsilon$ in Eqs. (\ref{dp}), (\ref{de}), and (\ref{moddisp})
may be related
such that $\eta_i \approx 0$ in Eq.~(\ref{eta})); (2) The time scale at
which quantum fluctuations of energy-momentum occur is relatively
short
\footnote{Unfortunately, these two scenarios also preclude the
possibility that energy-momentum uncertainties are the origin of the
threshold anomalies discussed above} (compared to the relevant interaction 
or decay times);
(3) Both ``systematic'' and ``non-systematic'' effects of quantum gravity
are present, but the ``systematic'' effects are large enough to overwhelm the
``non-systematic'' effects.\footnote{Some pundits may even entertain the
possibility that
energy-momentum fluctuations are negligible (in contrast to spacetime
fluctuations), for in that case, the kinematics that allows matter
instability is no longer operative.}

\bigskip

\section{\bf Coda: Other suggestions to probe Planck-scale physics}

In the preceding sections we have discussed several ways to probe
Planck-scale physics experimentally.  They include\\
1. looking for energy dependence of speed of light in timing arrival
of high energy $\gamma$-rays with the {\it same} as well as
{\it different} energies (section V);\\
2. looking for quantum foam-induced phase incoherence of light from
extragalactic sources (section VI);\\
3. using interferometers to detect displacement noise due to spacetime
fluctuations (section VII);\\
4. examining the UHECR and multi-TeV $\gamma$-ray events and explaining
the threshold anomalies (section VIII).\\
We have also suggested using black holes to probe Planck-scale 
physics\footnote{On the other hand, black hole physics and the associated
Hawking-Unruh effect may be probed experimentally via extremely violent
acceleration provided by a standing-wave of intense lasers\cite{chen}.}
theoretically (section III).  In principle, these are all potential
tests of quantum-gravity-induced uncertainties.  But, in practice, these
tests are either difficult or almost impossible (as seen in the preceding
sections).  Improved techniques and new ideas will be the key to 
test such ``{\it nonsystematic}'' quantum-gravity effects (in the 
terminology of Ref.\cite{ACNvD}).

There are other suggestions to probe Planck-scale physics that have
appeared in the literature.  Many of them provide constraints on the modified
dispersion relation Eq.~(\ref{systmod}), i.e., they are proposals to test some
of the purported ``{\it systematic}'' quantum-gravity effects.
In the following we list some of them:\\
1. Neutral kaon decay\cite{kaon}:
Laboratory experiments may probe possible quantum
nature of spacetime and possible CPT-violating effects induced by quantum
gravity in the neutral kaon system.\\
2. Clock comparison experiments and experiments with spin-polarized
matter\cite{myersetcetc}:
These experiments have been used to put stringent bounds on violations
of Lorentz symmetry and dispersion relations for nucleons, electrons,
photons and light quarks.\\
3. Vacuum Cherenkov effects\cite{majorJLM}:
Absence of such radiation and photon decay
can be used to constrain potential corrections to Lorentz
invariance.\\
4. Suppression of pion decay at high energy\cite{3expts}:
Experimental data on the
longitudinal development of the air showers produced by ultra-high energy
hadronic primaries appear to require that ultra-high energy neutral pions
are more stable than low energy pions as if the phase space for decay into
two photons is reduced at high energy on account of the modified dispersion
relation.\\
5. Synchrontron radiation in the Crab nebula\cite{JacobspanishAC}:
It has been argued that
observations of such radiation put extraordinarily stringent bounds on the
modifications of the dispersion relations for photons and electrons, but
some of the assumptions that go into the analysis have been
challenged.\\
6. Birefringence effects\cite{gleiser}:
Evidence of quantum gravity produced birefringence
can be searched for by analyzing polarized light from distant
sources.\\
7. High energy $\gamma$ rays from GRB\cite{Ellis,AEMNS,EllisPLB}:
By searching for different arrival
times for photons in different energy ranges (recall Eq.~(\ref{tdiff})), 
e.g., pulses emitted by
GRBs in different energy channels, one can test Poincare
symmetries and the dispersion relation Eq.~(\ref{systmod}).

Nowadays no review on Planck-scale physics is complete
without mentioning the so-called
``doubly special relativities''\cite{dsr} or ``modified special
relativities,''
proposed recently as possible Planck-scale
modifications of the ordinary Lorentz group.  They are related
to ``$\kappa$-deformed Lorentz symmetry''\cite{kappa}, and
provide non-linear realizations of Lorentz symmetry with two
fundamental invariants: the speed of light and a length scale usually
taken to be the Planck length.  Their geometries are non-commutative,
giving rise to deformed commutative relations, and generalized
uncertainty principles.  A related proposal\cite{Mendes}
based on the requirement of algebraic stability,
yields a class of relativistic quantum algebras
characterized by both the speed of light and a fundamental length
which can be taken to be the Planck length.

We conclude with a few remarks on two applications of 
Planck-scale physics to cosmology.  (See also Appendix B.)
Planck-scale physics and cosmology are linked
by the big bang theory.
Due to the redshifting that occurred during inflation,
wavelengths which correspond to cosmological lengths in the present era were
smaller than the Planck length during the early stages of inflation.  Thus
Planck-scale physics probably played a crucial role in the generation of
quantum modes in inflation.\cite{BrKeKi}  The effects of these modes might
be imprinted in the pattern of cosmological fluctuations we see in the
cosmic mircrowave background and the large-scale structure today.

Transplanckian physics may also account for the dark energy observed in
the present unverse.  By employing a nonlinear dispersion relation to
model the transplanckian regime, one can get ultralow frequencies at
very high momenta (or very short distances).\cite{Mersini}
It has been argued that
the ultralow energy modes are still frozen today by the expansion of
the universe.  Their energy provides a good candidate for the dark energy
which powers the accelerating expansion of the universe in the present era.

\bigskip

\begin{center}
{\bf Acknowledgments}
\end{center}

I am indebted to G. Amelino-Camelia and H. van Dam for very useful
discussions.  Most of what I know about Planck-scale physics I have
learned from them.  I also thank
A. Ashtekar, W. Christiansen, P. Frampton, C. Fuchs,
B.L. Hu, C.S. Lam, S. Lloyd, M. Ozawa,
N. Sasakura, S.H.Tye, W. Unruh, and R. Weiss for useful
discussions and constructive criticism.  Finally I thank K.K. Phua for
encouraging me to write this Brief Review, and
A. Giugni and L. Ng for their
help in preparing this manuscript.  This work was supported
in part by the U.S. Department of Energy and the Bahnson Fund of
the University of North Carolina.\\

\bigskip

\begin{center}
{\bf Appendix A: Dirac equation on a Planck-scale spacetime lattice}
\end{center}

If spacetime is discrete at the Planck scale, then fundamental equations
applicable
at short distances must be written in the form of difference equations
rather than differential equations.  In this appendix we present a form
of Dirac equation on a ``cubic" ($1 + 1$)-dimensional spacetime lattice.
We assume spacetime to be filled with a cubic lattice with lattice constant
$\Delta z = c \Delta t = l_P$.
Our approach\cite{geom} bears some resemblance to Feynman's\cite{Feyn} and
is based on Dirac's
observation\cite{Dirac} that the instantaneous velocity operators of the
spin-$\frac {1}{2}$
particle (hereafter called by the generic name ``the electron'') have
eigenvalues $\pm c$ and that they anticommute.
We assume that the electron of mass $m$ moves with the speed of light
($c = 1$ hereafter) from
one lattice
site to a neighboring (spatially left or right) site with time $t$ always
increasing on the ``cubic''
spacetime (z,t) lattice.  The wavefunction has two components
\begin{equation}
\psi (z,t) = \left( \begin{array}{c}
                     \psi_{+}(z,t) \\ \psi_{-}(z,t)
                     \end{array}  \right),
\label{psi1}
\end{equation}
where $\psi_{+}$ denotes the component arriving from the event $(z - l_P,
t - l_P)$ while $\psi_{-}$ means arriving from $(z + l_P, t - l_P)$.

Next we assume that, at the lattice site $(z, t)$, the arriving components
are partially turned around by a unitary matrix:
\begin{equation}
\left( \begin{array}{c}
           \psi_{+} (z, t) \\ \psi_{-} (z, t)
          \end{array}   \right)
  = \mathcal{F}  \left( \begin{array}{c}
                              \psi_{+} (z - l_P, t - l_P) \\
                              \psi_{-} (z + l_P, t - l_P)
                             \end{array}    \right),
\label{disDir}
\end{equation}
with the ``flip operator'' $\mathcal{F}$ defined by
\begin{equation}
\mathcal{F} \equiv e^{-iFm l_P}.
\label{flip}
\end{equation}
Here $F$ is a hermitian $2 \times 2$ matrix which we 
give the most obvious form
\begin{equation}
F = \sigma_{1} = \left( \begin{array}{cc}
                                         0 & 1 \\
                                         1 & 0
                                        \end{array}  \right),
\label{sigma1}
\end{equation}
with $\sigma_{1}$ being the first Pauli matrix.  We will show that,
in the limit $l_P \rightarrow 0$, the lowest nontrivial term of Eq.
(\ref{disDir}) yields the Dirac equation.  We first write
\begin{equation}
\left( \begin{array}{c}
           \psi_{+}(z - l_P, t - l_P) \\
           \psi_{-}(z + l_P, t - l_P)
          \end{array}   \right)
        = \mathcal{T} \psi(z, t),
\label{jump1}
\end{equation}
with the ``transfer'' operator $\mathcal{T}$ given by
\begin{equation}
\mathcal{T} = e^{- l_P \left(\frac {\partial}{\partial t} +\sigma_3
\frac
{\partial}{\partial z}\right)},
\label{jump2}
\end{equation}
where $\sigma_3 \equiv diag (1, -1)$ is the third Pauli matrix.
Then the difference equation Eq. (\ref{disDir}) takes the form
\begin{equation}
\psi (z, t) = \mathcal{F} \mathcal{T} \psi (z, t).
\label{allDir}
\end{equation}

The difference equation becomes a differential equation if we limit
ourselves to the zeroth order (given by the identity
$\psi (z, t) = \psi (z,t)$) and the first order term in $l_P$.
The first order equation is
\begin{equation}
i \frac {\partial}{\partial t} \psi (z, t) = m \sigma_1 \psi (z, t)
-i \sigma_3 \frac {\partial}{\partial z} \psi (z, t),
\label{Dirac}
\end{equation}
which is the Dirac equation in $(1 + 1)$ dimensions!

In the scenario we have proposed, the electron travels between
lattice sites
with the speed of light.  We\cite{geom} speculate that
only at the lattice sites does the electron
``feel'' its mass and perhaps also the presence of all external
fields.
But if gravitational
interactions also take place mainly at the lattice sites, does that mean
spacetime vertices somehow play an important role in
concentrating curvature?  And if so, how is this description of geometry and
topology related to the
Regge calculus\cite{Regge}, for example?  These problems deserve further
invesigation.

\bigskip
\begin{center}
{\bf Appendix B: $\Lambda$ as an interplay between Planck- and
Hubble-scale physics}
\end{center}

Recent astrophysical observations indicate that the
cosmological constant is probably small but nonzero; it is positive,
giving rise to cosmic repulsion, and is the source of the dark energy,
accounting for about $70\%$ of the total cosmic (critical) energy density.
In this appendix, we examine a scenario which suggests that the dark
energy may come about as a result of the interplay between Planck-scale
and Hubble-scale physics.  The idea makes crucial use of the theory of
unimodular gravity\cite{Bij,PRL}, which, for the purpose of this Brief
Report, can be regarded as the ordinary theory of gravity except for the
way the cosmological constant $\Lambda$ arises in the theory.

We use the version of unimodular gravity given by the
Henneaux and Teitelboim action\cite{ht}
\begin{equation}
S_{unimod} = - \frac{1}{16 \pi G} \int [ \sqrt{g} (R + 2 \Lambda) - 2
\Lambda
\partial_\mu {\mathcal T}^\mu](d^3x)dt.
\label{HT}
\end{equation}
One of its equations of motion is $\sqrt{g} = \partial_\mu
\mathcal{T}^\mu$,
the generalized unimodular condition, with $g$ given in terms of the
auxiliary field $\mathcal{T}^{\mu}$.  Note that, in this theory,
$\Lambda / G$ plays the role of
``momentum'' conjugate to the ``coordinate'' $\int d^3x {\mathcal T}_0$
which can
be identified, with the aid of the generalized unimodular condition, as
the spacetime volume $V$.  Hence $\Lambda /G$ and $V$ are
conjugate to each other.  It follows that
their fluctuations obey a Heisenberg-type
quantum uncertainty principle,
\begin{equation}
\delta V \! \delta \Lambda/G \sim 1.
\label{heisenb}
\end{equation}

Next we borrow an argument due to Sorkin\cite{sorkin},
drawn from the causal-set theory, which
stipulates that continous geometries in classical gravity should be
replaced by ``causal-sets'', the discrete substratum of spacetime.
In the framework of the causal-set theory, the
fluctuation in the number of elements $N$ making up the set is of the
Poisson type, i.e., $\delta N \sim \sqrt{N}$.  For a causal set, the
spacetime volume $V$ becomes $l_P^4 N$.  It follows that
\begin{equation}
\delta V \sim l_P^4\delta N \sim l_P^4 \sqrt{N}
\sim l_P^2\sqrt{V} = G \sqrt{V}.
\label{poisson}
\end{equation}
Putting Eqs. (\ref{heisenb}) and (\ref{poisson}) together yields a minimum
uncertainty in
$\Lambda$\ given by $\delta \Lambda \sim V^{-1/2}$.
By following an argument due to Baum\cite{Baum} and Hawking\cite{Hawk}, it
has been plausibly argued\cite{PRL} that, in the
framework of unimodular gravity,
$\Lambda$ vanishes to the lowest order of
approximation (in the ``late'' Universe)
and that it is positive if it is not zero.  So we conclude
that $\Lambda$ is positive and it fluctuates about zero
with a magnitude\cite{wuchen} of
$V^{-1/2} \sim R_H^{-2}$, where $R_H$ is the Hubble radius of the Universe,
contributing an energy density $\rho$ given by:
\begin{equation}
\rho \! \sim \! + \frac{1}{l_P^2 R_H^2},
\label{wow}
\end{equation}
which is of the order of the critical density as observed!  The appearance
of both the Planck length (the smallest length scale) and the Hubble radius
(the largest observable scale) in $\rho$ seems to suggest that
the dark energy is due to an interplay between ultraviolet and infrared
physics.\\

\end{document}